\documentclass{PoS}

\title{INTEGRAL and the unified model of AGN}

\ShortTitle{INTEGRAL and the unified model of AGN}

\author{\speaker{Claudio Ricci}\\%
       ISDC Data Centre for Astrophysics \\
       Geneva Observatory, University of Geneva\\
       E-mail: \email{claudio.ricci@unige.ch}}

\author{R. Walter and T. J.-L. Courvoisier\\
        ISDC Data Centre for Astrophysics \\
        Geneva Observatory, University of Geneva \\}
\abstract{According to the unified model, the differences observed in different classes of AGN are due to anisotropic obscuration. This implies that in the hard X-rays, at energies where the radiation can pierce the obscuring material, the average spectral characteristics of different classes of AGN should be the same. Here we present a study of the 17--250 keV {\it INTEGRAL} IBIS/ISGRI average spectra of Seyfert galaxies, aimed to constrain similarities and possible differences between different classes of AGN. 
The sample we consists of all the 172 Seyfert galaxies detected by {\it INTEGRAL} IBIS/ISGRI at z<0.2 during its first 7 years of observations. Of these, 45 are Seyfert\,1s, 31 Seyfert\,1.5s, 67 Compton-thin Seyfert\,2s, 11 Narrow Line Seyfert\,1s, 11 Compton-thick Seyfert\,2s, and 9 LINERs.
Comparing the average hard X-ray spectra of the different classes, we found that, while the spectra of Seyfert 1s, Seyfert 1.5 and NLS1s are consistent, Compton-thin Seyfert\,2 show a clear excess in the 20--60\,keV band over the spectrum of Seyfert\,1s and Seyfert\,1.5s. This excess might be due to an intrinsically stronger reflection component or to a clumpy structure of the absorber in Compton-thin Seyfert\,2s.

}

\FullConference{8th INTEGRAL Workshop ``The Restless Gamma-ray Universe"\\
		 September 27-30 2010\\
		 Dublin Castle, Dublin, Ireland}

\begin{document}

\section{Introduction}
Active Galactic Nuclei (AGN) are thought to be powered by accretion onto a super massive black hole (Rees 1984). Seyfert galaxies host AGN, and are commonly divided in Seyfert\,1s and Seyfert\,2s. While Seyfert\,1 galaxies show both broad (FWHM $\geq 1,000 \rm\,km\,s^{-1}$) and narrow (FWHM $\sim 300-1,000 \rm\,km\,s^{-1}$) optical emission lines, Seyfert\,2s show only narrow lines. Seyfert\,1.5s show an optical spectrum intermediate between those of Seyfert\,1s and Seyfert\,2s.
Spectropolarimetric studies of Seyfert\,2s (Miller \& Antonucci 1983) showed the existence of broad emission lines also in some Seyfert\,2s, which led to the formation of the unified model (UM) of AGN (Antonucci \& Miller 1985). According to this model the same engine is at work in all kind of Seyfert galaxies, and differences between Seyfert\,1s and Seyfert\,2s are due to the presence of an anisotropic absorber, which covers the broad line region (BLR) in Seyfert\,2s. This absorber is often associated to a molecular torus. Several other lines of evidence support the UM, like the deficit of ionizing photons in Seyfert\,2 galaxies, which indicates that the ionizing source is hidden from direct view (Schmitt \& Kinney 1996), and the fact that the Narrow Line Regions (NLRs) are seen with conical shapes in Seyfert\,2 galaxies and halo-like shapes in Seyfert\,1s (Pogge 1988).
However, in the last years evidence of differences between Seyfert\,1s and Seyfert\,2s has been discovered. Seyfert\,1s with significant absorption have been found (e.g., Cappi et al. 2006), along with Seyfert\,2s without X-ray absorption (e.g., Bianchi et al. 2008), and Seyferts with significantly clumpy absorbers (e.g., Ricci et al. 2010).
Furthermore spectropolarimetric surveys indicate that only $\sim 30-50\%$ of the Seyfert\,2s show polarized broad lines (PBL), which might imply that not all of them harbor hidden BLRs (Tran 2001,2003). 
The hard X-rays are particularly well suited to test the unified model, as at these energies the photons are unaffected by photoelectric absorption and it is possible to have a direct view of the X-ray source, at least up to values of the column densities $N_{\rm \,H}\simeq \sigma_{T}^{-1}$. For $N_{\rm \,H}> \sigma_{T}^{-1}$ Compton scattering starts in fact to play an important role, and only $\lesssim 50\%$ of the incoming photons are detected. Thus, considering only Compton-thin objects, one would expect to detect consistent average spectra for different classes of AGN.

%Possible differences have been found between the hard X-ray spectra of different types of Seyfert galaxies. Early studies using data obtained by {\it Ginga} and {\it Compton Gamma-Ray Observatory}/OSSE (Zdziarski et al. 1995) showed that the spectra of Seyfert\,2 are harder than those of Seyfert\,1. This was later confirmed by Gondek et al. (1996) using combined {\it EXOSAT}, {\it Ginga}, {\it HEAO-1} and {\it Compton Gamma-Ray Observatory}/OSSE, and by Done \& Smith (1996), using only {\it Ginga} data. More recently, a study of the average 3--200 keV {\it BeppoSAX} spectra of 23 Seyfert galaxies (Malizia et al. 2003) showed a substantially harder emission for Seyfert\,2 than for Seyfert\,1. Using a bigger sample of 45 Seyfert galaxies, Deluit \& Courvoisier (2003) found that the average {\it BeppoSAX}/PDS spectrum of Seyfert\,1 differs from that of Seyfert\,2, the first requiring the presence of a high energy cutoff, absent in the second, and possibly of a smaller reflection component.  Ajello et al. (2008) using {\it Swift}/BAT data also found evidence that on average the spectra of Seyfert\,2 are harder than those of Seyfert\,1.

\section{Sample and data analysis}
\begin{figure}[t!]
\centering
\includegraphics[width=7cm]{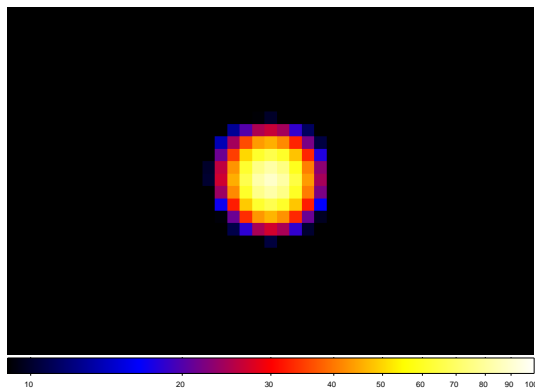}
\caption{Composite of the central part of the 17-80~keV mosaic image obtained for Seyfert\,1s.}
\label{fig:ima_1}
\end{figure}

The sample we used consists of all the 172 Seyfert galaxies detected by {\it INTEGRAL} IBIS/ISGRI (Ubertini et al. 2003) at z<0.2 during its first 7 years of observations. Of these, 45 are Seyfert\,1s, 31 Seyfert\,1.5s, 67 Compton-thin Seyfert\,2s, 11 Narrow Line Seyfert\,1s (NLS1s), 11 Compton-thick (CT) Seyfert\,2s and 9 LINERs. The optical classifications were taken from Veron-Cetty \& Veron (2010).
To extract the average spectra of the different AGN samples we followed the procedure adopted by Walter \& Cabral (2009). We created 500$\times$500-pixels mosaic images modifying the coordinate system of each individual image, setting the coordinates of each source of the sample to an arbitrary fixed position ($\alpha$=0, $\delta$=0). The geometry of the image was also modified to have consistent PSF whatever the position of the source in the field of view (FOV). These mosaic images provide a stack of all the selected IBIS/ISGRI data for each considered sample. Individual sky images for each pointing were produced in a broad energy band (i.e. 17--250~keV), and divided in 10 bins. The spectra were extracted from the mosaics using {\tt mosaic\_spec}. In figure\,\ref{fig:ima_1} we show, as an example, a composite of the central part of the 17-80~keV mosaic image obtained for Seyfert\,1s.

\section{Absorption in the X-rays}
Both photoelectric absorption and Compton scattering must be taken into account when modeling the effects of absorption in the X-rays.
The photoelectric cross section $\sigma_{ph}$ has a strong dependence on the energy, whereas Compton scattering depends on the Thomson cross section $\sigma_{\rm \,T}$, and it is constant with the energy, at least up to the Klein-Nishina decline. The ratio between Klein-Nishina and Thomson cross sections $\sigma_{\rm \,K-\,N}$/$\sigma_{\rm \,T}$ starts to diverge from the unity at $E \simeq 20$ keV, and it is of $\simeq 0.6$ at 200 keV. Compton processes become significant for column densities higher than $\sigma_{\rm \,T}^{-1}=1.5\times 10^{24} \rm \,cm^{-2}$.
The cumulative effect of the two cross sections is given by:
\begin{equation}\label{abs}
M(E)=e^{-\sigma_{ph}(E)N_{\rm \,H}}\times e^{-\sigma_{\rm \,T}N_{\rm \,H}}.
\end{equation}

\begin{figure}[h]
\centering
\includegraphics[width=9cm]{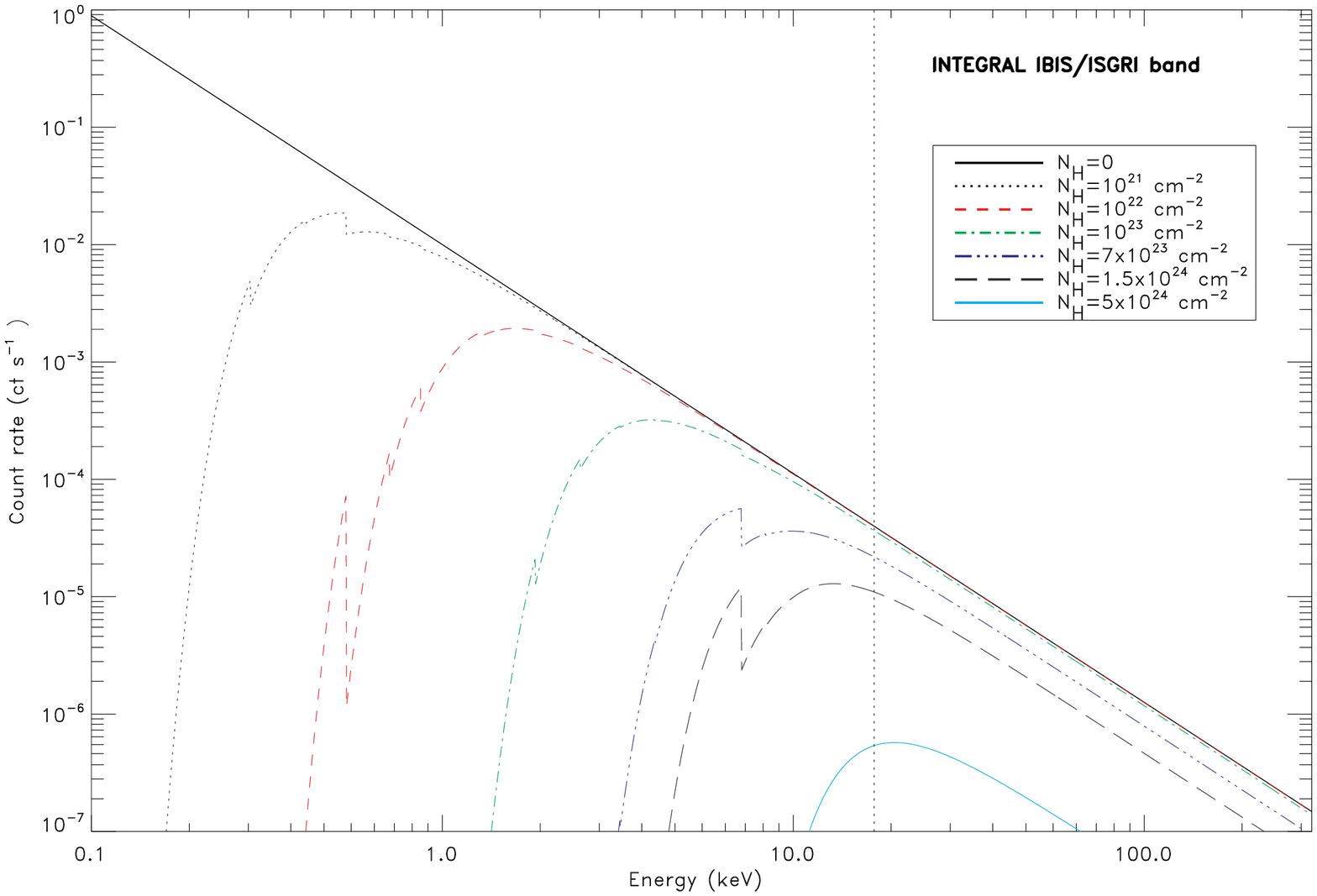}
\caption{Effect of photoelectric absorption and Compton scattering on a power law model with a photon index of $\Gamma=1.95$ in the X-rays.}
\label{fig:xrayabs}
\end{figure}

\begin{figure}[h]
\centering
\includegraphics[width=9cm]{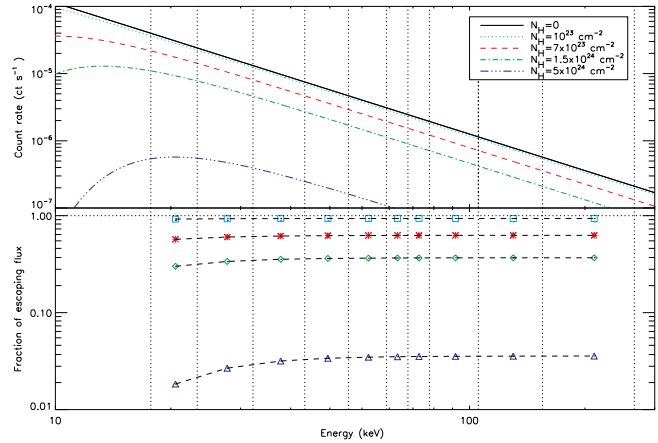}
\caption{Effect of photoelectric absorption and Compton scattering on a power law spectrum ($\Gamma=1.95$) in the hard X-rays. The upper panel shows the observed spectrum for different values of the hydrogen column density $N_{\rm \,H}$, and the lower panel shows the fraction of escaping flux in the 10 energy bins used in this work. In the lower panel squares represent the case $N_{\rm \,H}=10^{23} \rm \,cm^{-2}$, stars $N_{\rm \,H}=7\times 10^{23} \rm \,cm^{-2}$, diamonds $N_{\rm \,H}=1.5\times 10^{24} \rm \,cm^{-2}$ and triangles $N_{\rm \,H}=5\times10^{24} \rm \,cm^{-2}$.}
\label{fig:escapingflux}
\end{figure}

In Fig.\,\ref{fig:xrayabs} we show the effect of absorption (considering both photoelectric absorption and Compton scattering, as reported in Eq. \ref{abs}) on a power law model with a photon index of $\Gamma=1.95$ in the 0.1--300 keV energy range for different values of $N_{\rm\,H}$. In Fig.\,\ref{fig:escapingflux} we show the fraction of escaping flux for different values of $N_{\rm\,H}$ in the energy band we used. About 60\% of the original emission is unabsorbed for $N_{\rm \,H}=7\times10^{23}\rm \,cm^{-2}$, and we used this value as a threshold between Compton-thin and Compton-thick sources. In our final sample we considered only AGN with a value of the column density $N_{\rm \,H} \leq 7\times10^{23}\rm \,cm^{-2}$. The values of the column densities were taken from soft X-ray ($E<10\rm\,keV$) observations. We excluded from our sample the sources for which no value of $N_{\rm \,H}$ was available in the literature.
%We considered different values of the column density $N_{\rm \,H}$, and calculated the fraction of escaping flux in the energy bins we use. One can see that for column densities of the order of $\sigma_{\rm \,T}^{-1}$ only about 30\% of the flux is able to escape in the first bin. 
%In the following we will use solar abundances and the photoelectric cross section of Morrison and McCammon (1983). The choice of $\sigma_{ph}$ does not affect significantly our results, in fact using the more recent cross section of Verner et al. (1996), we would obtain a difference of only the $\sim\,3\%$ at $\simeq\,20$\,keV for $N_{\rm \,H}=7\times10^{23}\rm \,cm^{-2}$.

\section{Model-independent spectral analysis}
\begin{figure*}[h!]
\centering
\begin{minipage}[!b]{.48\textwidth}
\centering
\includegraphics[width=7.5cm]{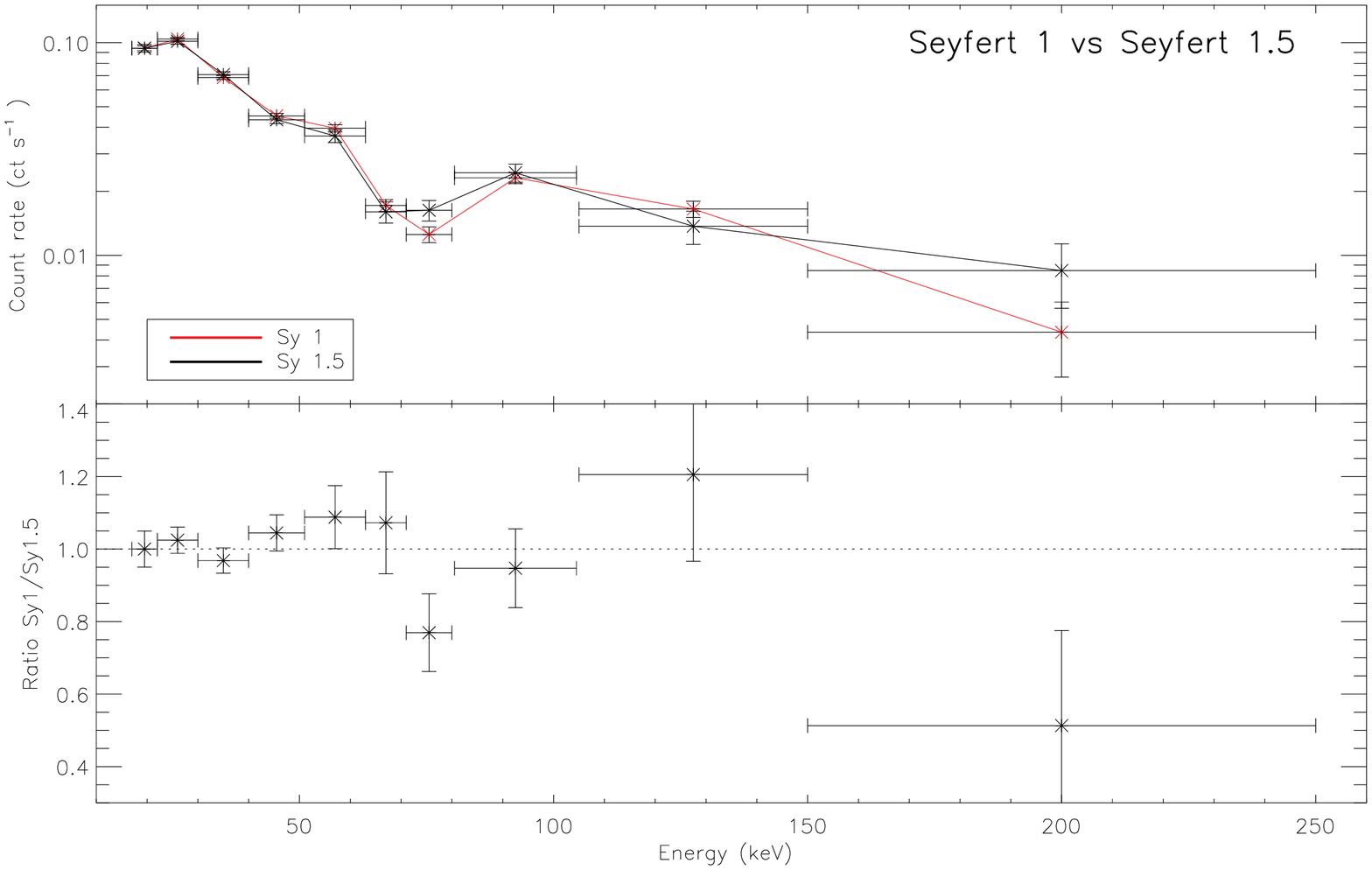} \end{minipage}
\begin{minipage}[!b]{.48\textwidth}
\centering
\includegraphics[width=7.5cm]{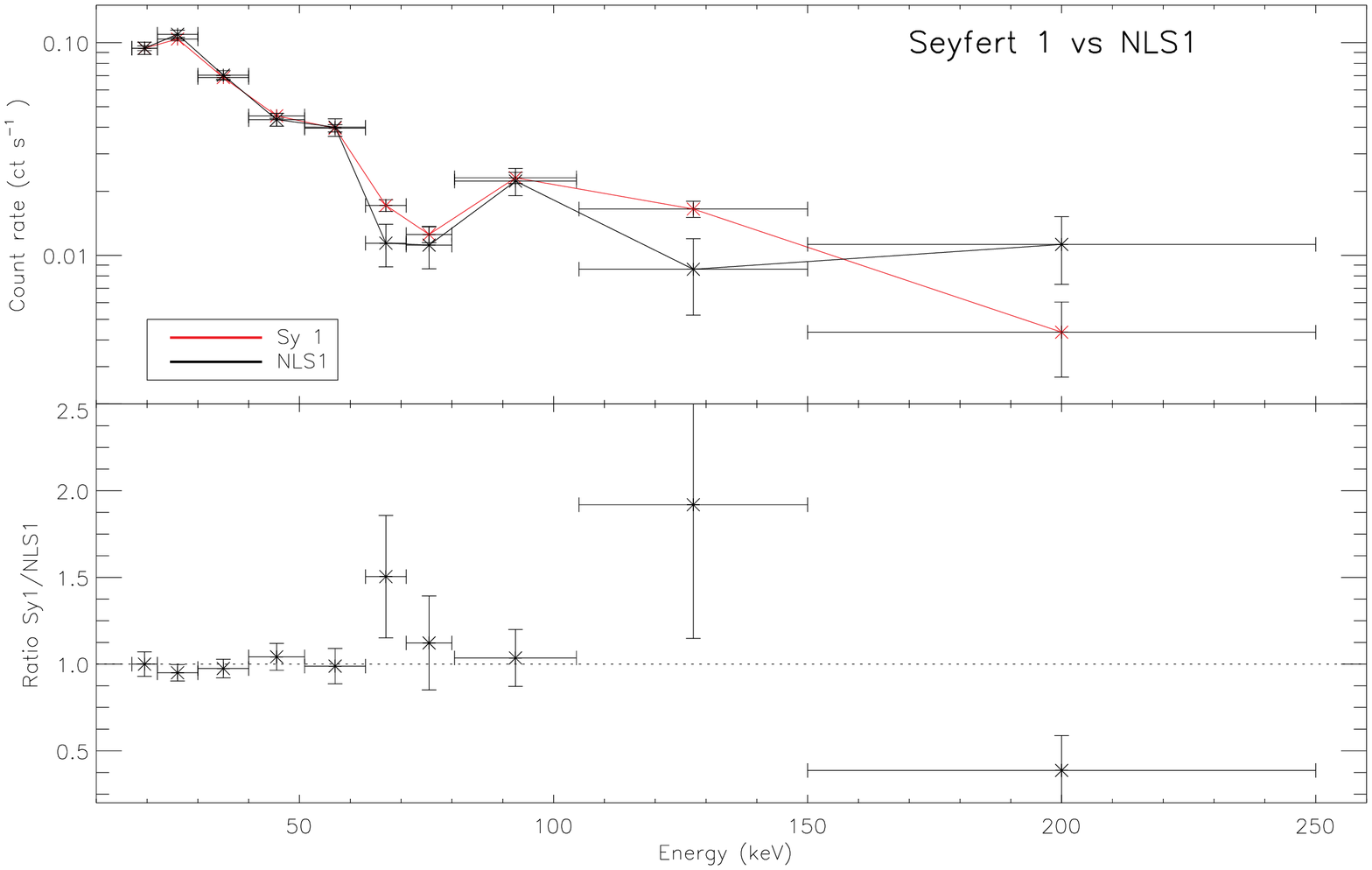}\end{minipage}
\hspace{0.05cm}
 \begin{minipage}[t]{1\textwidth}
  \caption{Spectra ({\it upper panels}) and ratios ({\it lower panels}) between the normalized spectra of Seyfert\,1s and Seyfert\,1.5s ({\it left}), Seyfert\,1s and NLS1s ({\it right}).}
\label{fig:modIndependent}
 \end{minipage}

\end{figure*}

The hard X-ray spectrum of AGN can be well described by a power law continuum with an exponential cut-off, and a reflection hump peaking at E~$\simeq 30$ keV (Magdziarz \& Zdziarski 1995).
One of the drawbacks of a model dependent analysis is the parameter degeneracy. This might not allow to well constrain several parameters at the same time. An alternative method to characterize differences and similarities between different classes of Seyfert galaxies, independently of their average flux, is through a model independent approach. This has been done by normalizing the flux of the different spectra at the first bin (i.e. 17--22 keV), and then calculating their ratio.
In Fig.\,\ref{fig:modIndependent} we show the spectra and ratios obtained comparing the average hard X-ray spectrum of Seyfert\,1s to those of Seyfert\,1.5s and NLS1s. As it can be seen from the figure the average spectra of both Seyfert\,1.5s and NLS1s are consistent with that of Seyfert\,1s. NLS1s are known to have steeper soft X-ray spectra than Seyfert\,1 and Seyfert\,1.5 galaxies. The fact that our sample of NLS1s is hard X-ray selected probably introduces a bias, being the objects with a flatter spectrum more easily detected than those with a steeper spectrum. Malizia et al. (2008) studying a smaller sample of NLS1s found that their hard X-ray spectrum is on average steeper than that of Seyfert\,1s. The difference between our results and theirs is probably due to the different sample, lower exposures and smaller energy ranges used in their work.

\begin{figure}[h]
\centering
\includegraphics[width=9cm]{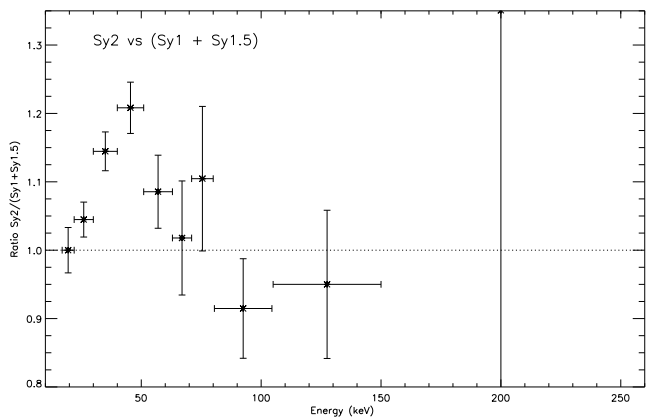}
\caption{Ratio between the spectra of Compton-thin Seyfert~2s and the one of Seyfert~1s and Seyfert~1.5s.}
\label{fig:merged}
\end{figure}

Given the similarity of the spectra of Seyfert\,1s and Seyfert\,1.5s, we re-ran the analysis considering all the objects of the two samples, obtaining an average spectrum of Seyfert\,1s and Seyfert\,1.5s.
We compared this spectrum to that of Compton-thin Seyfert\,2s, obtaining a result remarkably different from what we had for the other classes. As it can be seen from Fig.\,\ref{fig:merged}, the average hard X-ray spectrum of Compton-thin Seyfert\,2s is significantly harder in the 22--60 keV band, showing a bump which might be linked to a stronger reflection hump. At energies higher than 60\,keV the ratio of the two spectra is consistent with the unity, which implies that, although showing a stronger reflection component, Compton-thin Sy\,2s have the same continuum of Sy\,1s and Sy\,1.5s.

\section{The unified model of AGN}
The continuum emission of different classes of Seyfert galaxies has on average the same characteristics, which confirms the idea that different classes of objects have the same engine, as predicted by the zero-th order unified model of AGN. The stronger reflection observed in the spectrum of Compton-thin Seyfert\,2s cannot be explained in terms of absorption, as the chosen threshold of $N_{\rm\,H}$ should guarantee a transmission dominated spectrum. Moreover considering that the average column density of our sample of Seyfert\,2s is $N_{\rm\,H}=1.5\times10^{23}\rm\,cm^{-2}$, one would expect only $\simeq 10\%$ of the flux to be Compton scattered. The excess in the 20--60 keV band might be related either to an intrinsic stronger reflection component of Seyfert\,2s, or to the presence of clumps in the torus, as predicted by Elitzur \& Shlosman (2006).

\section{Conclusions}
We presented the study of the average {\it INTEGRAL} IBIS/ISGRI hard X-ray spectra of different classes of Seyfert galaxies. We found that, as predicted by the UM, the average spectra of Seyfert\,1s and Seyfert\,1.5s are consistent in the 17--250 keV band. NLS1s also show a spectrum similar to that of Seyfert\,1s and Seyfert\,1.5s, probably due to the hard X-ray selected nature of our sample. The average emission of Compton-thin Sy\,2s is harder than that of Sy\,1s and Sy\,1.5s in the 20--60 keV energy band, which might be due either to an intrinsically stronger Compton reflection or to a clumpy torus. The UM is confirmed on the zero-th order, although an homogenous anisotropic absorber alone cannot easily explain the differences between the spectrum of Seyfert\,2s and that of Seyfert\,1s and Seyfert\,1.5s.

\end{document}